\documentclass[english]{article}
\usepackage[T1]{fontenc}
\usepackage[latin9]{inputenc}
\usepackage{amsmath}
\usepackage{amssymb}
\usepackage{babel}

\begin{document}

\title{Discussion of ``Functional Models for Time-Varying Random Objects'' \\by Dubey and M\"uller}
\author{Dino Sejdinovic}
\date{\emph{Department of Statistics, University of Oxford}}

\maketitle
I congratulate the authors of \cite{DubeyMuller2020} on several substantial conceptual and theoretical
contributions which promise to lead to a widely applicable methodology. 
One of them is a new association measure between paired
random objects in a metric space, termed \emph{metric covariance.}
I will focus my discussion on this notion and on its relationship
with other similar concepts which have previously appeared in the
literature, including \emph{distance covariance} \cite{Szekely2007,Szekely2009,Lyons2013},
as well as its generalisations which rely on the formalism of reproducing
kernel Hilbert spaces (RKHS) \cite{Sejdinovic2013}. 

If $(\Omega,d)$ is a metric space such that $d^2$ is of negative type, then metric covariance (henceforth, mCov)
takes the form
\[
\text{cov}_{\Omega}(X,Y)=\frac{1}{4}\mathbb{E}_{XY}\mathbb{E}_{X'Y'}\left\{ d^{2}(X,Y')+d^{2}(X',Y)-2d^{2}(X,Y)\right\} .
\]
Negative type of $d^2$ implies that one can find a Hilbert space $\mathcal{H}$
and a \emph{feature map} $\phi:\Omega\to\mathcal{H}$ such that 
\[
d^{2}(X,Y)=\Vert\phi(X)-\phi(Y)\Vert_{\mathcal{H}}^{2},
\]
and hence
\begin{eqnarray*}
 \text{cov}_\Omega(X,Y) & = & \mathbb E_{XY}\left\langle \phi(X)-\mathbb E_X \phi(X), \phi(Y)-\mathbb E_Y \phi(Y)\right\rangle_{\mathcal H}\\
      {}& = & \mathbb E_{XY} \left\langle \phi(X), \phi(Y)\right\rangle_{\mathcal H} - \mathbb E_{XY'} \left\langle \phi(X), \phi(Y')\right\rangle_{\mathcal H},
\end{eqnarray*}
corresponding to the discrepancy between expected inner-products of features of $X$ and $Y$
under the joint and under the product of the marginals, measuring if $X$ and $Y$ are on average more similar (as measured by feature maps) in the coupled or in the uncoupled regime. Importantly, mCov can take both positive and negative values.

On the other hand, distance covariance (dCov) takes the following form
\begin{eqnarray*}
\Xi(X,Y) & = & \mathbb{E}_{XY}\mathbb{E}_{X'Y'}\rho_{\mathcal X}\left(X,X'\right)\rho_{\mathcal Y}\left(Y,Y'\right)\\
 &  & \qquad+\,\mathbb{E}_{X}\mathbb{E}_{X'}\rho_{\mathcal X}\left(X,X'\right)\mathbb{E}_{Y}\mathbb{E}_{Y'}\rho_{\mathcal Y}\left(Y,Y'\right)\\
 &  & \qquad\qquad-\,2\mathbb{E}_{XY}\left[\mathbb{E}_{X'}\rho_{\mathcal X}\left(X,X'\right)\mathbb{E}_{Y'}\rho_{\mathcal Y}\left(Y,Y'\right)\right],
\end{eqnarray*}
where $(\mathcal X,\rho_{\mathcal X})$, $(\mathcal Y,\rho_{\mathcal Y})$ are two semimetric spaces of negative type (we allow random objects $X$ and $Y$ to take values in different domains) and semimetrics $\rho_{\mathcal X}$ and $\rho_{\mathcal Y}$ take the role of $d^2$. This expression appears less intuitive and without an obvious link to mCov.

An alternative way to introduce dCov, however, is through the lens of RKHSs. Consider random objects
$X$ and $Y$ taking values on $\mathcal{X}$ and
$\mathcal{Y}$ respectively, and any two positive definite kernel functions $k:\mathcal{X}\times\mathcal{X}\to\mathbb{R}$
and $l:\mathcal{Y}\times\mathcal{Y}\to\mathbb{R}$ which are associated to RKHSs $\mathcal H_k$ and $\mathcal H_l$. Define the \emph{cross-covariance operator} $\Sigma_{YX}:\mathcal H_k \to \mathcal H_l$ such that
 \begin{equation*}
  \langle g,\Sigma_{YX}f\rangle_{\mathcal H_l} = \text{cov}\left[ f(X),g(Y)\right],\qquad \forall f\in\mathcal H_k, g\in\mathcal H_l.
 \end{equation*} %to considering embeddings
%of probability measures into RKHS, with $\mu_{k}:P\mapsto\int_{\mathcal{X}}k\left(\cdot,x\right)dP(x)$
%(as a Bochner integral) where $P$ is a Borel probability measure
%on a topological space $\mathcal{X}$ and $k:\mathcal{X}\times\mathcal{X}\to\mathbb{R}$
%is a positive definite kernel function associated to an RKHS $\mathcal{H}_{k}$.
The Hilbert-Schmidt Independence Criterion (HSIC), a notion (up to a constant factor) equivalent to dCov \cite{Sejdinovic2013} is given by

\begin{equation}
\Xi\left(X,Y\right)=\Vert \Sigma_{YX} \Vert_{HS}^2,    
\end{equation}
i.e. it is simply the squared Hilbert-Schmidt norm of feature-space cross-covariance. For a broad class of choices of $k$, $l$ -- in particular, \emph{characteristic kernels} \cite{Sriperumbudur2011} -- HSIC fully characterizes statistical dependence. These kernels include a widely used Gaussian kernel $k(x,x')=\exp(-\frac{1}{2\sigma^{2}}\left\Vert x-x'\right\Vert _{2}^{2})$ and the Mat\'{e}rn family.

%, for random objects
%$X$ and $Y$ taking values on topological spaces $\mathcal{X}$ and
%$\mathcal{Y}$ with respect to kernels $k:\mathcal{X}\times\mathcal{X}\to\mathbb{R}$
%and $l:\mathcal{Y}\times\mathcal{Y}\to\mathbb{R}$ is defined as the squared
%Hilbert space distance between the embeddings of the joint distribution
%of $X$ and $Y$ and the product of the marginals:
%\[
%\Xi_{k,l}\left(X,Y\right)=\left\Vert \mu_{k\otimes l}\left(P_{X,Y}\right)-\mu_{k}\left(P_{X}\right)\otimes\mu_{l}\left(P_{Y}\right)\right\Vert _{\mathcal{H}_{k\otimes l}}^{2}.
%\]
dCov can be recovered from HSIC by considering ``distance'' 
\begin{equation}
d_\mathcal X^{2}\left(x,x'\right)=k(x,x)+k\left(x',x'\right)-2k\left(x,x'\right)\label{eq: squared_distance}
\end{equation}
on $\mathcal{X}$ and similarly for $\mathcal{Y}$. Conversely, given any $d^2$ of negative type, we can construct the corresponding kernel
\begin{equation}
k\left(x,x'\right)=\frac{1}{2}\left(d^{2}\left(x,\omega\right)+d^{2}\left(x',\omega\right)-d^{2}\left(x,x'\right)\right)
\label{eq:kernel}
\end{equation}
where $\omega$ is an arbitrary anchor point.

% Alternative definition of HSIC hinges on isometry between $\mathcal{H}_{k\otimes l}$ and $HS\left(\mathcal{H}_{k},\mathcal{H}_{l}\right)$. If we define the cross-covariance operator $\Sigma_{YX}:\mathcal H_k \to \mathcal H_l$ such that
%  \begin{equation*}
%   \langle g,\Sigma_{YX}f\rangle_{\mathcal H_l} = \text{cov}\left[ f(X),g(Y)\right],\qquad \forall f\in\mathcal H_k, g\in\mathcal H_l,
%  \end{equation*}
% then $$\Xi\left(X,Y\right)=\Vert \Sigma_{YX} \Vert_{HS}^2,$$ i.e. it is simply the Hilbert-Schmidt norm of feature-space cross-covariance. For a broad class of choices of $k$, $l$ -- \emph{characteristic kernels} \cite{Sriperumbudur2011} -- HSIC fully characterizes statistical dependence. These kernels include a widely used Gaussian kernel $k(x,x')=\exp(-\frac{1}{2\sigma^{2}}\left\Vert x-x'\right\Vert _{2}^{2})$ and the Mat\'{e}rn family.

Is there also an RKHS interpretation of \emph{mCov}?
Recall that the domains of $X$ and $Y$ in this context coincide and are given by a metric space $(\Omega,d)$ with $d^2$ of negative type. We associate to it a positive-definite kernel in \eqref{eq:kernel} with RKHS $\mathcal{H}_{k}$ and define the cross-covariance operator $\Sigma_{YX}$. We claim that $\text{cov}_\Omega(X,Y)=\text{Tr}(\Sigma_{YX})$. Indeed,
\begin{eqnarray*}
\text{Tr}\left(\Sigma_{YX}\right) & = & \text{Tr}\left(\mathbb{E}_{XY}k\left(\cdot,X\right)\otimes k\left(\cdot,Y\right)-\mathbb{E}_{XY'}k\left(\cdot,X\right)\otimes k\left(\cdot,Y'\right)\right)\\
 & = & \mathbb{E}_{XY}\text{Tr}\left(k\left(\cdot,X\right)\otimes k\left(\cdot,Y\right)\right)-\mathbb{E}_{XY'}\text{Tr}\left(k\left(\cdot,X\right)\otimes k\left(\cdot,Y'\right)\right)\\
 & = & \mathbb{E}_{XY}\left\langle k(\cdot,X),k(\cdot,Y)\right\rangle _{\mathcal{H}_{k}}-\mathbb{E}_{XY'}\left\langle k(\cdot,X),k(\cdot,Y')\right\rangle _{\mathcal{H}_{k}}\\
 & = & \mathbb{E}_{XY}k(X,Y)-\mathbb{E}_{XY'}k(X,Y')\\
 & = & \frac{1}{2}\left(\mathbb{E}_{XY'}d^{2}(X,Y')-\mathbb{E}_{XY}d^{2}(X,Y)\right).
\end{eqnarray*}
Recall that HSIC/dCov can be understood as
$$\Xi(X,Y)=\Vert \Sigma_{YX} \Vert_{HS}^2=\text{Tr}(\Sigma_{YX}\Sigma_{XY}),$$
so indeed the two notions are closely related. To further interpret the connection, we can take a Mercer basis $\{\sqrt{\lambda_{j}}e_{j}\}_{j\in J}$ of $\mathcal{H}_{k}$.
Then

\[
\text{cov}_\Omega(X,Y)=\sum_{j\in J}\lambda_{j}\langle e_j,\Sigma_{YX}e_j\rangle_{\mathcal H_k}=\sum_{j\in J}\lambda_{j}\text{cov}\left[e_{j}\left(X\right),e_{j}\left(Y\right)\right],
\]
i.e. mCov considers how evaluations at the \emph{same} basis function covary and it can be zero if positive and negative covariances between basis function evaluations ``cancel out''. In contrast, HSIC/dCov considers covariances of all pairs of basis function evaluations:
\[
\Xi(X,Y)=\sum_{i\in J}\sum_{j\in J}\lambda_i\lambda_{j}\text{cov}\left[e_{i}\left(X\right),e_{j}\left(Y\right)\right]^2.
\]

We will now consider some cases where mCov is zero even though the variables are dependent. A straightforward example is to consider the case where there exists dependence between $X$ and $Y$ but their feature representations live in orthogonal subspaces, e.g. if we take a \emph{linear} kernel on $\mathbb R^2$ and $X=(Z,0)$, $Y=(0,Z)$. A perhaps more interesting example, also in $\mathbb R^2$, is as follows: take $Z\sim Bern\left(\frac{1}{2}\right)$, and
\[
X\sim\begin{cases}
\mathcal{N}\left(\left[-1,+1\right],\sigma^{2}I\right), & \text{if }Z=0,\\
\mathcal{N}\left(\left[+1,-1\right],\sigma^{2}I\right), & \text{if }Z=1,
\end{cases}\qquad Y\sim\begin{cases}
\mathcal{N}\left(\left[-1,-1\right],\sigma^{2}I\right), & \text{if }Z=0,\\
\mathcal{N}\left(\left[+1,+1\right],\sigma^{2}I\right), & \text{if }Z=1.
\end{cases}
\]
We have here coupled the ``mixing variable'' so that $X_{1}$ and $Y_{1}$
are positively correlated, whereas $X_{2}$ and $Y_{2}$ are negatively
correlated. It is readily shown however that $\left\Vert X-Y\right\Vert \overset{d}{=}\left\Vert X-Y'\right\Vert $.
Hence, mCov computed with
\emph{any radial kernel}, i.e. where $k(x,y)$ depends on $x$ and $y$ through $\Vert x-y \Vert$ only, which includes Gaussian and Mat\'ern families known to be characteristic, will not be able to detect such
dependence between $X$ and $Y$. To be able to detect dependence
we would require looking into individual dimensions, which may become
impractical for higher dimensional problems.

In summary, while the authors demonstrate that dCov/HSIC is not suitable for use in the developed framework of object functional principal component analysis, it is worth noting that mCov is a strictly weaker statistical dependence measure and it is possible that it misses certain types of multivariate associations. For a generic choice of metric, the corresponding feature map $\phi$ is defined implicitly and may not be straightforward to interpret while which forms of dependence are captured by mCov does depend on the form of $\phi$ and hence on the associated kernel $k$. Finally, we believe that the RKHS interpretation described here may give rise to different estimation methods of mCov and to its novel uses.

\bibliography{biblio}
\bibliographystyle{plain}

\end{document}